\documentclass[conference]{IEEEtran}

\usepackage{graphicx}
\usepackage[dvipsnames]{xcolor}
\usepackage{soul}
\usepackage{cite}
\usepackage{amsmath,amssymb,amsfonts}
\usepackage{algorithmic}
\usepackage{graphicx}
\usepackage{textcomp}
\usepackage{balance}
\usepackage[colorlinks=true,urlcolor=blue,citecolor=black,linkcolor=black]{hyperref}
\usepackage[utf8]{inputenc}
\usepackage[T1]{fontenc}
\usepackage{url}
\usepackage{wrapfig}
\usepackage{caption}
\usepackage{algorithm}
\usepackage{hhline}
\usepackage{fixltx2e}
\usepackage{tabularx}
\usepackage{booktabs}
\usepackage{colortbl}
\usepackage{multirow}
\usepackage{enumitem}
\usepackage{flushend} 
\usepackage{pifont} 
\usepackage[dvipsnames]{xcolor}
\usepackage{orcidlink}

\usepackage[scaled=0.86]{helvet}

\usepackage{framed}

\definecolor{light-gray}{gray}{0.9}
\usepackage[framemethod=TikZ]{mdframed}

\mdfdefinestyle{MyFrame}{%
    linecolor=black,
    outerlinewidth=0.15pt,
    roundcorner=3pt,
    innertopmargin=2pt,
    innerbottommargin=2pt,
    innerrightmargin=4pt,
    innerleftmargin=4pt,
    backgroundcolor=light-gray}
    
\mdfdefinestyle{MyFrameSimple}{%
    linecolor=black,
    outerlinewidth=0.15pt,
    roundcorner=3pt,
    innertopmargin=2pt,
    innerbottommargin=2pt,
    innerrightmargin=4pt,
    innerleftmargin=4pt}

\newcommand{\sectopic}[1]{\vspace*{0.1em}\par\noindent{\textit{\bfseries #1}}}

\newcommand {\StepOne}{\color{Red}\ding{202}\color{Black}}
\newcommand {\StepTwo}{\color{Red}\ding{203}\color{Black}}
\newcommand {\StepThree}{\color{Red}\ding{204}\color{Black}}
\newcommand {\StepFour}{\color{Red}\ding{204}\color{Black}}

\ifCLASSINFOpdf
\else
\fi

\begin{document}
\title{Enhancing Legal Compliance and Regulation Analysis with Large Language Models
\vspace*{-.1em}}

\author{Shabnam Hassani \orcidlink{0009-0008-3056-4073} \\University of Ottawa (Email: s.hassani@uottawa.ca) }


\maketitle
\pagenumbering{arabic}
\thispagestyle{plain}
\pagestyle{plain} 

\begin{abstract}
This research explores the application of Large Language Models (LLMs) for automating the extraction of requirement-related legal content in the food safety domain and checking legal compliance of regulatory artifacts. With Industry 4.0 revolutionizing the food industry and with the General Data Protection Regulation (GDPR) reshaping privacy policies and data processing agreements, there is a growing gap between regulatory analysis and recent technological advancements. This study aims to bridge this gap by leveraging LLMs, namely BERT and GPT models, to accurately classify legal provisions and automate compliance checks. Our findings demonstrate promising results, indicating LLMs' significant potential to enhance legal compliance and regulatory analysis efficiency, notably by reducing manual workload and improving accuracy within reasonable time and financial constraints.


\end{abstract}

\IEEEpeerreviewmaketitle

\section{Introduction}\label{sec:introduction}
The evolution of Industry 4.0 and  regulations like GDPR demand innovative approaches to ensure that food safety software systems and regulatory artifacts, such as Data Processing Agreement (DPAs), comply with legal standards.
Our review of the literature and existing methodologies reveals a gap in applying LLMs to domain-specific regulatory analysis. This study is propelled by the need to automate the interpretation and application of legal provisions in the food industry and data processing regulatory artifacts and is also motivated by industry partners. This research aims to develop an LLM-based methodology for classifying food safety and checking the compliance of regulatory provisions, seeking to outperform traditional methods in efficiency, accuracy, and cost.

My research is driven by two main objectives: to develop an automated methodology for classifying legal provisions pertaining to food safety, and to enhance the efficacy and interoperability of compliance checking of regulatory artifacts using LLMs. I propose that LLMs can significantly outperform existing methods in these tasks, thereby enhancing legal compliance and regulation analysis with improvements in terms of accuracy, time, and financial cost.

\subsection{Research Questions}
The main research questions (RQs) are: 

\textbf{RQ1: How accurate is the legal requirements classification approach?} RQ1 examines alternative realizations of our classification approach using different BERT variants and GPT-3.5-turbo. To answer RQ1, we report accuracy using the standard classification metrics.

\textbf{RQ2: How does the legal requirements classification approach fare against baselines?} To assess whether our approach offers benefits over simpler solutions, RQ2 compares our approach against two baselines: one based on LSTM and the other based on keyword search. To answer RQ2, we report accuracy using the standard classification metrics.

\textbf{RQ3: How do state-of-the-art LLMs, both open-source and closed-source, fare against one another in terms of accuracy for zero-shot learning and fine-tuning in regulatory compliance checking?}
This research question compares generative LLMs in terms of their ability to understand and apply regulatory requirements. To answer RQ3, we report accuracy using the standard classification metrics.

\textbf{RQ4: Compared to traditional sentence-level analysis, how does incorporating paragraph context and the textual specification of compliance rules enhance the performance of compliance checking?}
This research question aims to assess the enhancement in accuracy of compliance checking brought about by the integration of paragraph-level context and rules, as opposed to traditional single-input sentence-level methods. To answer RQ4, we report accuracy using the standard classification metrics.

\textbf{RQ5: What are the cost and time implications associated with our proposed approach?}
To be practical, our approach must generate predictions within a reasonable time. RQ5 provides a practical assessment of the approach's resource efficiency, considering time and costs in terms of computing and financial resources.

\textbf{Structure.} Section~\ref{sec:Back} presents background. Section~\ref{sec:related} compares the proposed method with related work. Section~\ref{sec:approach} describes our proposed method. Section~\ref{sec:evaluation} provides current evaluation results. Section~\ref{sec:analysis} lays out the future research and the timeline. 
\section{background} \label{sec:Back}
In this section, the necessary legal and technical background for my approach is provided.

\subsection{Food-safety Regulations}
To ensure that the food supply is safe, countries around the world have established regulatory frameworks to govern the production, distribution, and consumption of food products. 
Our primary focus is on the Safe Food for Canadians Regulations (SFCR).
SFCR consolidates several federal Canadian food regulations that apply, among other things, to the import, export, and inter-provincial trade of food, including ingredients. 
In many cases, SFCR defers the elaboration of requirements for specific food products to Food-Specific Requirements and Guidance (FSRG) regulations. 
We use content from FSRG regulations both in our qualitative analysis and in our evaluation. In addition to SFCR and FSRG regulations, we consider selected parts of the food-safety regulations by the US Food and Drug Administration (FDA) to examine the extent to which our classification solution is transferable to jurisdictions outside Canada.

\subsection{General Data Protection Regulation (GDPR)}
\label{sec:GDPR}
GDPR is a privacy and data protection law enacted by the European Union to regulate the processing of personal data and enhance individuals' control over their information.
We use Data Processing Agreements (DPAs), as defined by GDPR, to illustrate ideas and conduct evaluations. DPAs are binding agreements that specify the responsibilities and rights of controllers and processors. Most of the content in DPAs is software-related, making them of direct interest to the RE community~\cite{Amaral2023ML}. For GDPR compliance, DPAs must follow GDPR Article~28.


\section{Related Work} \label{sec:related}
Significant progress has been made in compliance checking of regulatory artifacts. Below we discuss limitations in existing research for compliance checking fall under two categories.

For our classification of requirements-related legal content from food safety regulations, we consider three prior lines of work: (1)~extracting metadata from legal texts, (2)~applying LLMs to natural-language requirements, and (3)~monitoring food safety through IoT. To our knowledge, no prior work exists on automated classification of requirements-related provisions in food-safety regulations.

\subsection{Related Work on Compliance Checking of Legal Content}
 
\subsubsection{Reliance on sentences as units of analysis}
Existing approaches mostly promote a sentence-by-sentence analysis of legal texts~\cite{Amaral2021AI,Amaral2023NLP,Amaral2023ML,ilyas2023multi}. We have observed three key issues caused by this decision that can significantly affect accuracy.
First, interpreting sentences often requires contextual understanding beyond the immediate sentence structure, as sentences can be related to previously mentioned categories or definitions.
Second, it is common for a legal concept to be defined, with subsequent sentences drawing upon the definition.
Third, legal texts are frequently interwoven with cross-references~\cite{Sannier2017Automated}. This practice adds complexity by requiring an understanding of the broader content across a collection of legal documents.

\subsubsection{Automation strategies lack justification for decisions and are either coarse-grained or entail significant manual effort to build}
Automation for legal texts employs one of the following three strategies or a combination thereof. We argue that these strategies not only fail to offer justification and rationale for decisions -- a capability now achievable with LLMs -- but they either are too coarse-grained, potentially compromising accuracy, or require considerable manual effort, making compliance automation costly to implement.

\textit{\textbf{Strategy~1}}: Extracting metadata or semantic frames and then executing predefined rules over the metadata/frames to check compliance such as~\cite{Amaral2021Model,Amaral2021AI, Amaral2023ML,Amaral2023NLP,xiang2023policychecker}.

\textit{\textbf{Strategy~2}}: Projecting regulatory texts into an embedding space and then measuring the similarity between these texts against (textual) compliance rules or a labelled group of regulatory provisions. An example of research employing this strategy is that of Amaral et al.~\cite{Amaral2021AI,Amaral2023ML}, who create sentence embeddings and utilize semantic similarity for completeness checking of privacy policies and DPAs.

\textit{\textbf{Strategy~3}}: Using encoder-only single-input transformer models, such as BERT~\cite{Devlin2018Bert}, to classify which compliance rules are satisfied by each legal text. An example of work applying this strategy is that of Ilyas et al.~\cite{ilyas2023multi}, who use BERT to construct a multi-label classification pipeline for identifying sentences in DPAs that meet specific compliance rules.

\subsection{Related Work on the Classification of Requirements-related Legal Content}

\subsubsection{Metadata Extraction from Legal Texts}
Legal documents are typically lengthy and complex, making it difficult for individuals to locate relevant information quickly and accurately. This has led to the development of automated approaches for information extraction from legal texts such as~\cite{Breaux2006Towards, Breaux2008Analyzing,Breaux2013Preserving,Bhatia2018Semantic}. These works, despite being amongst the first to focus on information extraction from legal texts, do not use ML techniques and are limited to privacy policies.


Studies such as~\cite{Amaral2021AI,Amaral2023NLP,Amaral2023ML} differ from ours in both domain and the utilization of feature-based learning as opposed to LLM-based approaches.

Some studies link privacy policies to software code\cite{Fan2020Empirical,Hamdani2021Combined,Xie2022Scrutinizing}. In contrast, our approach focuses on food-safety regulations. We use BERT variants and GPT for classification and information extraction.

Zeni et al.~\cite{Zeni2015Gaiust,Zeni2016Building} and Sleimi et al.~\cite{Sleimi2018Automated} extract semantic metadata from legal texts using NLP and semantic-web techniques, but their focus and domain differ from ours. Abualhaija et al.~\cite{Abualhaija2022Automated} employ LLMs for question answering over GDPR; they have different analytical goals and inputs from ours.

\subsubsection{LLMs and Natural-language (NL) Requirements}
Several studies use LLMs, particularly BERT, for analyzing NL requirements, with tasks including classifying non-functional requirements~\cite{Hey2020NoRBERT,Chatterjee2021Pipeline,Alhoshan2022Zero-Shot}, classifying and summarizing contractual obligations~\cite{Sainani2020Extracting,jain2023transformer}, detecting requirements smells~\cite{Habib2021Detecting}, classifying security requirements~\cite{Varenov2021Security}, classifying user feedback~\cite{Mekala2021Classifying}, classifying requirements dependencies~\cite{Deshpande2021BERT}, detecting causality~\cite{Fischbach2021Automatic}, classifying and clustering coreferences~\cite{Wang2020Deep,Wang2022Detecting}, transforming NL requirements into formal specifications~\cite{Nayak2022Req2Spec}, predicting issue links~\cite{Luders2022Automated}, classifying issue sentences~\cite{Mehder2022Classification}, identifying similar requirement~\cite{Abbas2022Relationship}, checking  completeness~\cite{Luitel2023Using}, and generating elicitation-interview scripts~\cite{Gorer23Generating}. These studies showcase the versatility of LLMs in requirements-analysis tasks. We have benefited from the best practices in the above-cited strands of work; however, our analytical objectives set us apart.

\subsubsection{Food-monitoring Systems}
Food monitoring is a crucial area of research for improving food safety and enhancing supply-chain efficiency. Bouzembrak et al.~\cite{Bouzembrak2019Internet} have conducted a systematic literature review to examine the potential of IoT for food safety and quality monitoring, as well as food traceability and supply chain. They observe that most existing IoT research aimed at these objectives focuses on measurement solutions using sensors and communication technologies. These solutions are technology-driven and not directly linked or traceable to regulations. Our qualitative study highlights, from a regulatory perspective, the importance of all the core measurement types identified in the literature. The close alignment between the measurements implied by food-safety regulations and the existing measurement technologies confirms the validity of our interpretation of systems/software relevance in the context of food safety.
\section{Proposed Method} \label{sec:approach}
We conduct a comprehensive qualitative analysis of food safety regulations followed by utilizing LLMs, specifically BERT and GPT-3.5. We also conduct an analysis of compliance checking of regulatory artifacts with regulations followed by utilizing LLMs specifically GPT-3.5, GPT-4, Mixtral, and BERT. The methodology is further described in detail. 

\subsection{Automated Classification of Legal Provisions}

\begin{figure}[!b]
\vspace*{-1em}
\centering    \includegraphics[width=\linewidth]{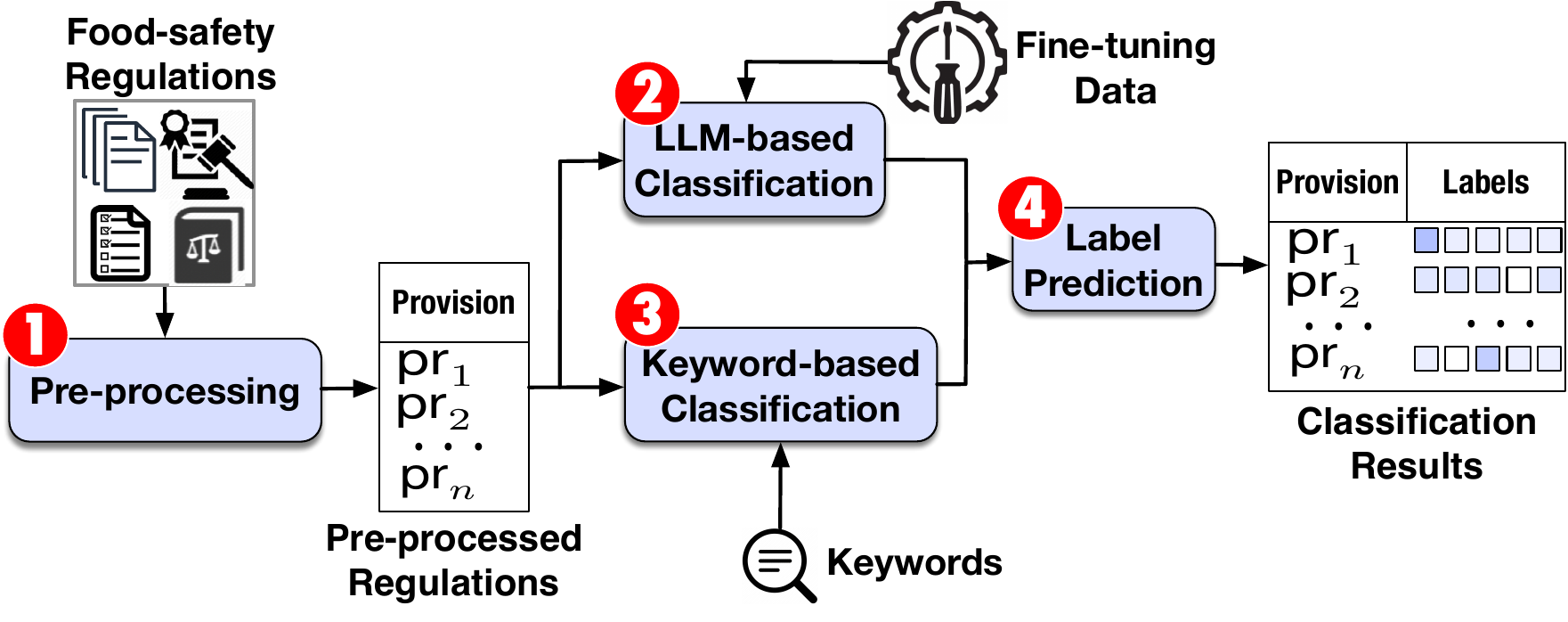}
\vspace*{-1em}
\caption{Overview of our automated classification approach.}\label{fig:approach1}
\end{figure}

Figure~\ref{fig:approach1} presents an overview of our approach for the automated classification of requirements-related legal content, specifically instantiated for requirements-related concepts in food-safety regulations.
The approach takes as input the textual content of the regulations being examined and produces as output labels for each provision in the input. The approach treats each sentence as one provision. Stated otherwise, our \emph{unit of analysis} is a sentence since larger units like paragraphs and sections may not be precise enough for this task.
Our approach has four steps: (\StepOne) Pre-processing, (\StepTwo) LLM-based Classification, (\StepThree) Keyword-based Classification, and (\StepFour) Label Prediction (Fig~\ref{fig:approach1}).

\sectopic{Step~1)~Pre-processing.} In this step, we split the input into sentences. 
Legal documents frequently use lists. To ensure preservation of context during sentence splitting, we follow Sleimi et al.~\cite{Sleimi2018Automated} and add the header in each list, known as a list-item prefix, to each itemized/enumerated item.

\sectopic{Step~2)~LLM-based Classification.} Using an LLM fine-tuned on labelled data from our qualitative study, we classify provisions with occurrences of concepts from our conceptual model. In our evaluation, we examine BERT and GPT as alternative LLMs. For GPT, the pre-processed regulations from Step~1 are embedded into prompts. 

\sectopic{Step~3)~Keyword-based Classification.}\label{subsec:keyword-based}
Our qualitative study did not yield a sufficiently large number of examples for a few concepts: \emph{Colour}, \emph{Firmness}, \emph{Pathogen}, and \emph{Water Content}. As previous studies on automated classification of legal texts point out~\cite{Sleimi2018Automated,Torre2020Ai,Amaral2021AI}, some concepts may turn out to be too scarce for learning-based techniques, thereby warranting the use of keywords for classification. In the third step of our approach, we conduct a keyword lookup for the four scarce concepts mentioned above. The keywords associated with each concept were derived during our qualitative study. If a provision $P$ contains one or more of the keywords associated with a concept $M$, we label $P$ with $M$. 


\sectopic{Step~4)~Label Prediction.} This step combines the labels computed by the LLM-based classifiers (Step~2) and those computed by the keyword-based classifiers (Step~3) to produce the final label recommendations for each provision. 

\subsection{Automated Compliance Checking}
\begin{figure}
\centering    \includegraphics[width=.8\linewidth]{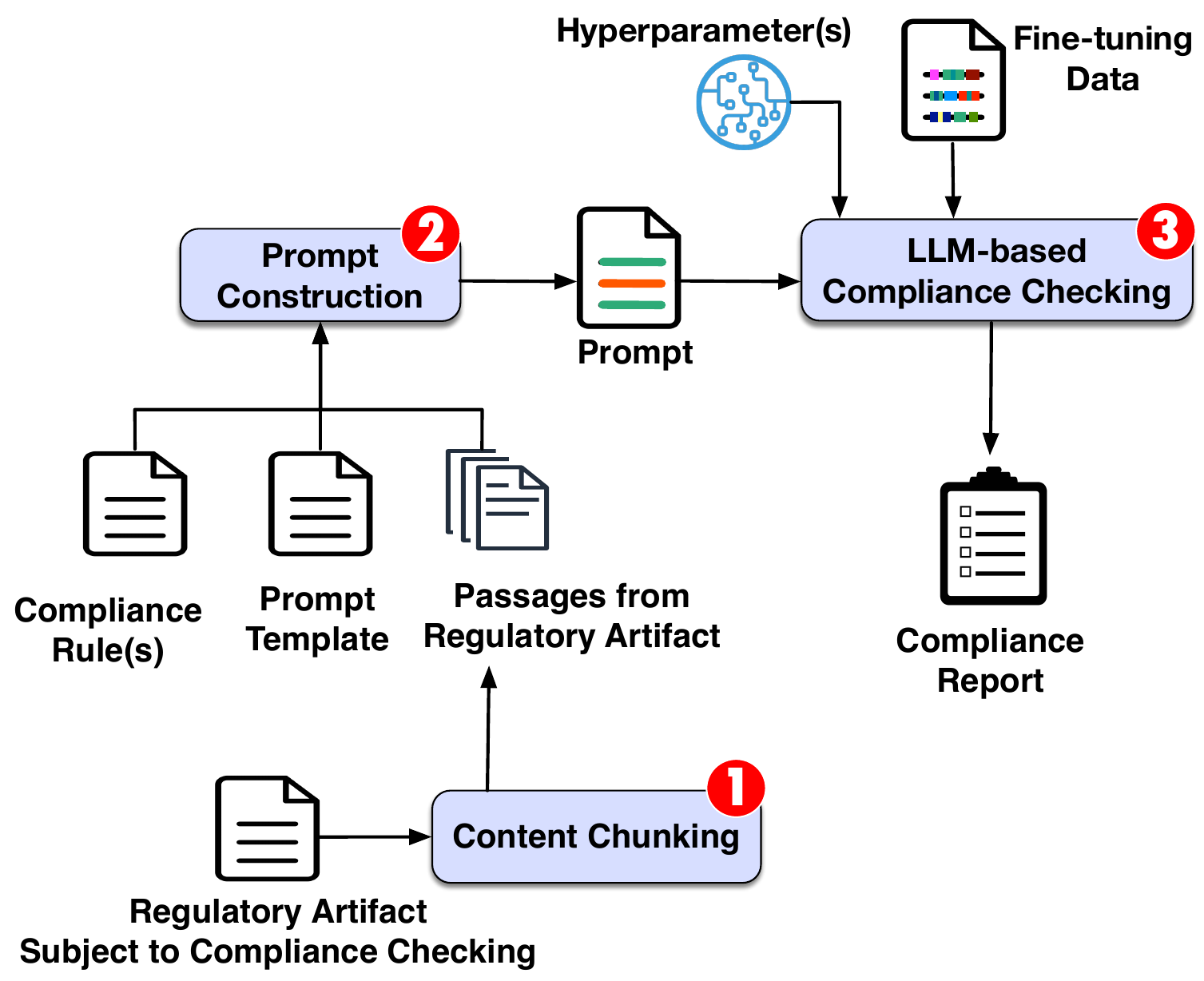}
\caption{Overview of our automated compliance checking approach.}
\label{fig:approach2}
\vspace*{-1em}
\end{figure}

Figure~\ref{fig:approach2} presents an overview of our approach for automated compliance checking, which systematically employs LLMs to assess regulatory artifacts against specific regulations.
The approach consists of three steps. The first step (\StepOne) is to create passages from the regulatory artifact that the user wants to verify for compliance, e.g., a DPA. The second step (\StepTwo) is to generate a prompt for LLMs. This step takes as input the passages generated in the first step, along with the compliance rules to be verified (e.g., from the GDPR), and a configurable prompt template. The third step (\StepThree) is to present the prompt generated in the second step to LLMs to draw inferences about compliance and non-compliance. The output of the approach is a compliance report, accompanied by an explanation and justification for each determination.

\sectopic{Step~1)~Content Chunking.}
The regulatory artifact subject to compliance checking is fed to the pipeline. The pipeline segments the regulatory artifact into manageable content chunks,
primarily paragraphs (Fig.~\ref{fig:approach2}, Step~\StepOne).
This process ensures that the chunks, once incorporated into the overall prompt, will fit within the token limit of the LLM. 
If a full paragraph were to exceed the token limit, it would need to be either truncated or summarized. However, in our investigation of LLMs, we did not encounter such situations due to their reasonably large token limit.
The output of this step is a set of passages each within the LLM's token limit, ready for prompt construction.

\sectopic{Step~2)~Prompt Construction.}
Our approach constructs tailored prompts to guide the LLMs in eliciting rule-specific responses, based on the input compliance rules. 
Prompts, designated to extract both the rule applicability and the LLMs' explanation and justification, consist of three parts: \emph{Prompt Template}, \emph{Compliance Rules}, and \emph{Passages from Regulatory Artifact} (Fig.~\ref{fig:approach2}, Step~\StepTwo). 

The prompt is presented in a structured chat format for the LLMs, specifying roles for system instructions, user inquiries, and the assistant's responses, following best practices where applicable~\cite{OpenAiFinetuning,Chatcompletion}.
The system role provides instructions for the model to follow (detailed in the Prompt Template). The user role presents input that model should respond to (Passages from Regulatory Artifact and Compliance Rule(s)). The assistant role represents the model's response to the user's input (Compliance Report).
\textit {Prompt Template: You are a legal expert trained to identify applicable \{Compliance Rules\} based on a given \{text\} within its specific \{context\}. When provided with the \{text\} and its \{context\}, your response should only include the rule identifier (e.g., 'R5') if applicable. If there is no direct connection to any Compliance Rule within the context provided, respond with 'R99'. Follow this format strictly. Then, provide your rationale for the decision. }

\sectopic{Step~3)~LLM-based Compliance Checking.}
Once tailored prompts are constructed, we employ zero-shot learning, where the model generates responses without prior training on similar tasks, based solely on the instructions provided in the prompt. In the future, this process may be enhanced with (optional) fine-tuning to further refine the models' responses to the specific language of DPAs. 
Subsequently, the generated responses are analyzed to determine the compliance of the DPA text with the GDPR, identifying areas of compliance and non-compliance in a comprehensive report (Fig.~\ref{fig:approach2}, Step~\StepThree).


\section{Current Empirical Results} \label{sec:evaluation}
We conducted (1)~a qualitative study that characterizes food-safety concepts in regulatory provisions impacting modern software-intensive food-safety systems, and (2)~an LLM-based approach that classifies the provisions of food-safety regulations based on their relevance to systems and software requirements. We conducted an extensive evaluation of our approach by instantiating it with both BERT and GPT-3.5. 

We evaluated our models using standard classification metrics, including \emph{Accuracy}, \emph{Precision}, \emph{Recall}, and \emph{F-score}, to classify legal provisions effectively. This evaluation facilitated a comprehensive comparison of LLM performances, identifying strengths and highlighting areas for enhancement. To account for variability, we conducted multiple experiments and reported the average results. Boxplots were generated to visually represent the performance outcomes of these experiments.
Our results indicate that accuracy is largely consistent across BERT (\emph{Precision} of 87\%, \emph{Recall} of 86\%, and \emph{F-score} of 87\%) and GPT-3.5 (\emph{Precision} of 89\%, \emph{Recall} of 83\%, and \emph{F-score} of 86\%), with BERT achieving slightly better overall results.

For our compliance checking of DPAs with GDPR, our early experiments with \emph{gpt-3.5-turbo-0125}\cite{OpenAi}, \emph{Mixtral-8x7B-Instruct-v0.1}\cite{Mixtral}, and \emph{gpt-4-0125-preview}~\cite{OpenAi} show promising improvements when transitioning from sentence-level to paragraph-level passages. On average, the improvements in \emph{Accuracy} obtained through using these three models are in the ranges of 33\% (from 30\% and 63\%), 35\% (from 33\% and 69\%), and 40\% (from 41\% and 81\%), respectively.


\section{Future Research}\label{sec:analysis}
My research aims to make a case for the need to reconsider current practices in legal compliance automation in light of recent advances in AI. Specifically, I argue that the enhanced capacity of modern LLMs to handle context is likely to induce a major shift in our treatment of textual legal artifacts. This shift will involve transitioning from analyzing smaller contexts, such as individual sentences and phrases, to considering larger volumes of content, such as paragraphs and beyond, as context. I posit that the larger context will be able to provide the prerequisite knowledge, including cross-referenced legal materials, to create a self-contained basis for accurate automated decision-making regarding compliance.


In the domain of food-safety regulations, my focus will extend to exploring concepts beyond North America, with a keen interest in the regulations of countries like India and the UK, which, while operating under English law, have distinct regulations.


In the domain of DPAs, my future work will focus on four main aspects: (1)~enriching the DPA dataset~\cite{Amaral2023ML} with paragraph-level annotations; (2)~conducting comprehensive empirical evaluations to validate the effectiveness of paragraph-level context in increasing LLM accuracy; (3)~benchmarking against prior BERT-based approaches, e.g.,~\cite{ilyas2023multi}, to showcase benefits; and (4)~involving legal experts to critically assess LLM outputs, especially concerning their explanations and justifications. A pertinent question arises on evaluating legal experts' verification of these justifications.

I foresee dedicating at least eight months to these mentioned activities, paralleling these efforts with the composition of a research paper for submission to RE 2025.

In addition, and time permitting, I plan to examine techniques like Retrieval Augmented Generation~\cite{gao2023retrieval} to facilitate question answering over domains such as DPAs and Privacy Policies. A preparatory period of four months will be reserved for this initiative, post-investigation and prior to full-scale implementation.
As I progress, I trust that the constructive insights garnered from our ongoing research, coupled with the sufficient time frame, will empower me to not only expand upon my thesis but also to finalize it punctually by the culmination of my fourth academic year.

\makeatletter
\newcommand{\myfontsize}{\@setfontsize\myfontsize{8.35}{8.25}}
\makeatother

\let\oldthebibliography\thebibliography
\renewcommand{\thebibliography}[1]{%
  \oldthebibliography{#1}%
  \myfontsize 
  \setlength{\itemsep}{0pt}%
}

\clearpage
\bibliographystyle{IEEEtran}
\bibliography{ref}
\end{document}